\documentclass{article}

\begin{document}

\title{Modern Mathematical Physics:\\	what it should be}

\author{L.~D.~Faddeev}


\maketitle

	When somebody asks me, what I do in science, I call myself
	a specialist in mathematical physics. As I have been there for more
than 40 years, I have some definite interpretation of this
	combination of words: ``mathematical physics." Cynics or purists
	can insist that this is neither mathematics nor physics,
	adding comments with a different degree of malice. Naturally, this
	calls for an answer, and in this short essay I want to explain 
	briefly my understanding of the subject. It can be considered
	as my contribution to the discussion about the origin and role 
	of mathematical physics and thus to be relevant for this volume.

	The matter is complicated by the fact that the term ``mathematical
	physics" (often abbreviated by MP in what follows) is used 
	in different 
	senses and can have rather different content. This content changes
	with time, place and person.

	I did not study properly the history of science; however, it is my
	impression that, in the beginning of the twentieth century, the term MP
was
	practically equivalent to the concept of theoretical physics.
	Not only Henri Poincar\'e, but also Albert Einstein, were called
	mathematical physicists. Newly established theoretical chairs were
called chairs of mathematical physics. It follows from the 
	documents in the archives of the Nobel Committee that MP had a
	right to appear both in the nominations and discussion of the
	candidates for the Nobel Prize in physics
\cite{Nagel}.
	Roughly speaking, the concept of MP covered theoretical papers
	where mathematical formulae were used.

	However, during an unprecedented bloom of theoretical physics
	in the 20s and 30s, an essential separation of the terms ``theoretical"
	and ``mathematical" occurred. For many people, MP was reduced to
	the important but auxiliary course ``Methods of Mathematical 
	Physics" including a set of useful mathematical tools. The 
	monograph of P.~Morse and H.~Feshbach
\cite{Morse}
	is a classical example of such a course, addressed to a wide 
	circle of physicists and engineers.

	On the other hand, MP in the mathematical interpretation appeared
	as a theory of partial differential equations and variational
	calculus. The monographs of R.~Courant and D.~Hilbert
\cite{Courant}
	and S.~Sobolev
\cite{Sobolev}
	are outstanding illustrations of this development. The theorems
	of existence and uniqueness based on the variational principles,
	a priori estimates, and imbedding theorems for functional spaces
	comprise the main content of this direction. As a student of 
	O.~Ladyzhenskaya, I was immersed in this subject since the 3rd
	year of my undergraduate studies at the Physics Department 
	of Leningrad
	University. My fellow student N.~Uraltseva now holds the chair of
MP exactly
	in this sense.

	MP in this context has as its source mainly geometry and such parts
	of classical mechanics as hydrodynamics and elasticity theory.
	Since the 60s a new impetus to MP in this sense was supplied by Quantum
	Theory. Here the main apparatus is functional analysis, 
 including the spectral theory of operators in Hilbert space, the
	mathematical theory of scattering and the theory of Lie groups and
	their representations. The main subject is the Schr\"{o}dinger
	operator.
	Though the methods and concrete content of this part of MP are
	essentially different from those of its classical counterpart, the
	methodological attitude is the same. One sees the quest for the
	rigorous mathematical theorems about results which are understood
	by physicists in their own way.

	I was born as a scientist exactly in this environment. I graduated
	from the unique chair of Mathematical Physics, established by
	V.I.~Smirnov at the Physics Department of Leningrad University
	already in the 30s. In his venture V.I.~Smirnov got support from
	V.~Fock, the world famous theoretical physicist with very wide
	mathematical interests. Originally this chair played the auxiliary
	role of being responsible for the mathematical courses for 
	physics students. However in 1955 it got 
	permission to supervise its own diploma projects, and I belonged
	to the very first group of students using this opportunity.
	As I already mentioned, O.A.~Ladyzhenskaya was our main professor.
	Although her own interests were mostly in nonlinear PDE and
	hydrodynamics, she decided to direct me to quantum theory. During
	last two years of undergraduate studies I was to read the
	monograph of K.O.~Friedrichs, ``Mathematical Aspects of Quantum
	Field Theory," and relate it to our group of 5 students and our
	professor on a special seminar. At the same time my student friends
	from the chair of Theoretical Physics were absorbed in reading
	the first monograph on Quantum Electrodynamics by A.~Ahieser
	and V.~Berestevsky. The difference in attitudes and language was
	striking and I was to become accustomed to both.

	After my graduation O.A.~Ladyzhenskaya remained my tutor
	but she left me free to choose research topics and literature
	to read. I read both mathematical papers (i.e.\ on direct and
	inverse scattering problems
	by I.M.~Gelfand and B.M.~Levitan, V.A.~Marchenko, M.G.~Krein, 
	A.Ya.~Povzner)
	and ``Physical Review" (i.e.\ on formal scattering theory by 
	M.~Gell-Mann, M.~Goldberger, J.~Schwinger
	and H.~Ekstein) as well.
	Papers by I.~Segal, L.~Van-Hove and R.~Haag added to my first
	impressions on Quantum Field Theory taken from K.~Friederichs.
	In the process of this self-education my own understanding of the
	nature and goals of MP gradually deviated from the prevailing
	views of the members of the V.~Smirnov chair.
	I decided that it is more challenging to do something which
	is not known to my colleagues from theoretical physics rather
	than supply theorems of substantiality. My first work on the
	inverse scattering problem especially for the many-dimensional
	Schr\"{o}dinger operator and that on the three body scattering
	problem confirm that I really tried to follow this line of 
	thought.

	This attitude became even firmer when I began to work on Quantum
	Field Theory in the middle of the 60s. As a result, my 
	understanding of the goal of MP drastically modified. I consider
	as the main goal of MP the use of mathematical intuition for the
	derivation of really new results in the fundamental physics.
	In this sense, MP and Theoretical Physics are competitors.
	Their goals in unraveling the laws of the structure of matter
	coincide. However, the methods and even the estimates of the
	importance of the results of work may differ quite significally.

	Here it is time to say in what sense I use the term ``fundamental
	physics." The adjective ``fundamental" has many possible
	interpretations when applied to the classification of science.
	In a wider sense it is used to characterize the research directed
	to unraveling new properties of physical systems. In the narrow
	sense it is kept only for the search for the basic laws that
	govern and explain these properties.

	Thus, all chemical properties can be derived from the
	Schr\"{o}dinger equation for a system of electrons and nuclei.
	Alternatively, we can say that the fundamental laws of chemistry
	in a narrow sense are already known. This, of course, does not
	deprive chemistry of the right to be called a fundamental
	science in a wide sense.

	The same can be said about classical mechanics and the quantum
	physics of condensed matter. Whereas the largest part of physical 
	research lies now in the latter, it is clear that all its successes
	including the theory of superconductivity and superfluidity,
	Bose-Einstein condensation and quantum Hall effect have
	a fundamental explanation in the nonrelativistic quantum theory
	of many body systems.

	An unfinished physical fundamental problem in a narrow sense is
	physics of elementary particles. This puts this part of physics
	into a special position. And it is here where modern MP has
	the most probable chances for a breakthrough.

	Indeed, until recent time, all physics developed along the 
	traditional
	circle: experiment --- theoretical interpretation ---
	new experiment.
	So the theory traditionally followed the experiment. This imposes a
	severe censorship on the theoretical work. Any idea, bright as it is,
	which is not supplied by the experimental knowledge at the time
	when it appeared is to be considered wrong and as such must be
	abandoned. Characteristically the role of censors 
	might be played by
	theoreticians themselves and the great L.~Landau and W.~Pauli were, as
	far as I can judge, the most severe ones. And, of course, they
	had very good reason.

	On the other hand, the development of mathematics, which is also
	to a great extent 
	influenced by applications, has nevertheless its internal
	logic. Ideas are judged not by their relevance but more by esthetic
	criteria. The totalitarianism of theoretical physics gives way to a
	kind of democracy in mathematics and 
	its inherent intuition. And exactly
	this freedom could be found useful for particle physics.
	This part of physics traditionally is based on the
	progress of accelerator techniques. The very high cost and
	restricted possibilities of the latter soon will become an 
	uncircumventable obstacle to further development. And it is
	here that mathematical intuition could give an adequate 
	alternative. This was already stressed by famous theoreticians with
	mathematical inclinations. Indeed, let me cite a paper
\cite{Dirac}
	by P.~Dirac from the early 30s:
\begin{quotation}
	The steady progress of physics requires for its theoretical
	formulation a mathematics that gets continually more advanced.
	This is only natural and to be expected. What, however,
	was not expected by the scientific workers of the last century
	was the particular form that the line of advancement of the
	mathematics would take, namely, it was expected that the
	mathematics would get more complicated, but would rest on
	a permanent basis of axioms and definitions, while actually
	the modern physical developments have required a mathematics
	that continually shifts its foundations and gets more abstract.
	Non-euclidean geometry and non-commutative algebra, which
	were at one time considered to be purely fictions of
	the mind and pastimes for logical thinkers, have now been found
	to be very necessary for the description of general facts of the
	physical world. It seems likely that this process of increasing
	abstraction will continue in the future and that advance in
	physics is to be associated with a continual modification
	and generalization of the axioms at the base of mathematics
	rather than with logical development of any one mathematical 
	scheme on a fixed foundation.

	There are at present fundamental problems in theoretical
	phy\-sics awaiting solution, 
    \emph{e.g.},
	the relativistic formulation of quantum mechanics and the nature
	of atomic nuclei (to be followed by more difficult ones such as
	the problem of life), the solution of which problems will
	presumably require a more drastic revision of our fundamental
	concepts than any that have gone before. Quite likely these
	changes will be so great that it will be beyond the power of
	human intelligence to get the necessary new ideas by direct
	attempts to formulate the experimental data in mathematical
	terms. The theoretical worker in the future will therefore have
	to proceed in a more inderect way.
	The most powerful method of advance that can be suggested
	at present is to employ all the resources of pure mathematics
	in attempts to perfect and generalise the mathematical formalism
	that forms the existing basis of theoretical physics, and
    \emph{after}
	each success in this direction, to try to interpret the new
	mathematical features in terms of physical entities.
\end{quotation}
	Similar views were expressed by C.N.~Yang. I did not find a compact
	citation, but all spirit of his commentaries to his own collection
	of papers 
\cite{Yang1983}
	shows this attitude. Also he used to tell this to me in private 
	discussions.

	I believe that the dramatic history of setting the gauge fields as a
	basic tool in the description of interactions in Quantum Field Theory
	gives a good illustration of the influence of mathematical intuition
	on the development of the fundamental physics. Gauge fields, or 
	Yang--Mills fields, were introduced to the wide audience of physicists
in
	1954 in a short paper by C.N.~Yang and R.~Mills 
\cite{Yang1954},
	dedicated to the generalization of the electromagnetic fields and the 
	corresponding principle of gauge invariance. The geometric sense of 
	this principle for the electromagnetic field was made clear as early 
	as in the late 20s due to the papers of V.~Fock
\cite{Fock}
	and H.~Weyl
\cite{Weyl}.
	They underlined the analogy of the gauge (or gradient in the 
	terminology of V.~Fock) invariance of the electrodynamics and the 
	equivalence principle of the Einstein theory of gravitation. The 
	gauge group in electrodynamics is commutative and corresponds 
	to the multiplication of the complex field (or wave function) of 
	the electrically charged particle by a phase factor depending on
	the space--time coordinates. 
	Einstein's theory of gravity provides an example of a
	much more sophisticated gauge group, namely the group of general
	coordinate transformation. Both H.~Weyl and V.~Fock were to use the
	language of the moving frame with spin connection, associated with 
	local Lorentz rotations. Thus the Lorentz group became the first
	nonabelian gauge group and one can see in 
\cite{Fock}
	essentially all formulas characteristics of nonabelian gauge
	fields. However, in contradistinction to the electromagnetic field, 
	the spin connection enters the description of the space-time and not
	the internal space of electric charge.

	In the middle of the 30s, after
	the discovery of the isotopic spin in nuclear physics, 
	and forming the Yukawa idea of the intermediate boson,
	O.~Klein tried to geometrise these objects. His proposal was
	based on his 5-dimensional picture. Proton and neutron (as well as
	electron and neutrino, there were no clear distinction between
	strong and weak interactions) were put together in an
	isovector and electromagnetic field and charged vector meson
	comprised a 
$ 2 \times 2 $
	matrix. However the noncommutative
$ SU(2) $
	gauge group was not mentioned.

	Klein's proposal was not received favorably and N.~Borh did not
	recommend him to publish a paper. So the idea remained only in
	the form of contribution to proceedings of Warsaw Conference
	``New Theories in Physics"
\cite{Klein}.

	The noncommutative group, acting in the internal space of
	charges, appeared for the first time in the paper
\cite{Yang1954}
	of C.N.~Yang and R.~Mills in 1954. There is no wonder
	that Yang received a cool reaction 
	when he 
	presented his work at Princeton in 1954. The dramatic account of 
	this event can be found in his commentaries
\cite{Yang1983}.
	Pauli was in the audience and immediately raised the question 
	about mass. 
	Indeed the gauge invariance forbids the introduction of mass
	to the vector charged fields and masslessness leads to the
	long range interaction, which contradicts the experiment.
	The only known massless particles (and accompaning long range
	interactions) are photon and graviton.
	It is evident from Yang's text, that Pauli was
	well acquainted with the differential geometry of nonabelian
	vector fields but his own censorship
	did not allow him to speak about them.
	As we know now, the boldness of Yang and his esthetic feeling
	finally were vindicated. And it can be rightly said, that C.N.~Yang
	proceeded according to mathematical intuition.
	
	In 1954 the paper of Yang and Mills did not move
	to the forefront of high energy theoretical physics. However, the
	idea of the charged space with noncommutative symmetry group acquired
	more and more popularity due to the increasing number of elementary
	particles and the search for the universal scheme of their 
	classification. And at that time the decisive role in the promotion
	of the Yang--Mills fields was also played by mathematical intuition.

	At the beginning of the 60s, R.~Feynman worked on the extension of 
	his own scheme of quantization of the electromagnetic field to the 
	gravitation theory of Einstein. A purely technical difficulty ---
	the abundance of the tensor indices --- made his work rather slow. 
	Following the advice of M.~Gell-Mann, he exercised first on the
	simpler case of the Yang--Mills fields. To his surprise, he found
	that a naive generalization of his diagrammatic rules designed for
	electrodynamics did not work 
	for the Yang-Mills field. The unitarity of the
    $ S $-matrix was broken. Feynman restored the unitarity in one
loop by
	reconstructing the full scattering amplitude from its imaginary
	part and found that the result can be interpreted as a subtraction
	of the contribution of some fictitious particle. However his technique
	became quite cumbersome beyond one loop. His approach
	was gradually developed by B.~De-Witt
\cite{DeWitt}.
	It must be stressed that
	the physical senselessness of the Yang--Mills field did not
	preclude Feynman from using it for mathematical construction.

	The work of Feynman 
\cite{Feynman}
	became one of the starting points for my work in Quantum Field Theory,
	which I began in the middle of the 60s together with Victor Popov. 
	Another point as important was the mathematical monograph by 
	A.~Lichnerowitz 
\cite{Lichnerowicz},
	dedicated to the theory of connections in vector bundles. From 
	Lichnerowitz's book it followed clearly that the Yang--Mills field
	has a definite geometric interpretation: it defines a connection
	in the vector bundle, the base being the space-time and the
	fiber the linear space of the representation of the compact group
	of charges. Thus, the Yang--Mills field finds its natural place among
	the fields of geometrical origin between the electromagnetic field
	(which is its particular example for the one-dimensional charge) 
	and Einstein's gravitation field, which deals with the tangent 
	bundle of the
	Riemannian space-time manifold. 

	It became clear to me that such a possibility cannot be missed and, 
	notwithstanding the unsolved problem of zero mass, one must actively 
	tackle the problem of the correct quantization of the Yang--Mills
	field.

	The geometric origin of the Yang--Mills field gave a natural way
	to resolve the difficulties with the diagrammatic rules.
	The formulation of the quantum theory in terms of Feynman's
	functional integral happened to be most appropriate from the
	technical point of view. Indeed, to take into account the gauge
	equivalence principle one has to integrate over the classes of
	gauge equivalent fields rather than over every individual
	configuration. As soon as this idea is understood, the technical
	realization is rather straightforward. As a result V.~Popov
	and I came out at the end of 1966 with a set of rules valid for
	all orders of perturbation theory. The fictitious particles appeared
	as auxiliary variables giving the integral representation for
	the nontrivial determinant entering the measure over the set
	of gauge orbits.

	Correct diagrammatic rules of quantization of the Yang-Mills field, 
	obtained by V.~Popov and me in 1966--1967 
\cite{Faddeev,Popov},
	did not attract immediate the attention of physicists. Moreover, the
time
	when our work was done was not favorable for it. Quantum Field 
	Theory was virtually forbidden, especially in the Soviet Union,
	due to the influence of Landau. ``The Hamiltonian is dead" --- this
	phrase from his paper
\cite{Landau}, 
	dedicated to the anniversary of W.~Pauli ---
	shows the extreme of Landau's attitude. The reason was quite solid,
	it was based not on experiment, but on the investigation of the 
	effects of renormalization, which led Landau and his coworkers to
	believe that the renormalized physical coupling constant is inevitably
	zero for all possible local interactions. So there was no way for 
	Victor Popov and me to publish an extended article in a major
	Soviet journal. We opted for the short communication in
	``Physics Letters" and were happy to be able to publish the
	full version in the preprint series of newly opened Kiev Institute
	of Theoretical Physics. This preprint was finally translated into
	English by B.~Lee as a Fermilab preprint in 1972, and from the 
	preface to the translation it follows that it was known in the West
	already in 1968.

	A decisive role in the successful promotion of our diagrammatic
	rules into physics was played by the works of G.~'t~Hooft
\cite{Hooft1971},
	dedicated to the Yang--Mills field interacting with the Higgs field 
	(and which ultimately led to a Nobel Prize for him in 1999) and the 
	discovery of dimensional transmutation (the term of S.~Coleman
\cite{Coleman}). 
	The problem of mass was solved in the first case via the 
	spontaneous symmetry breaking. The second development was based on
	asymptotic freedom. There exists a vast literature dedicated to the 
	history of this dramatic development. I refer to the recent papers 
	of G.~'t~Hooft 
\cite{Hooft1998}
	and D.~Gross
\cite{Gross},
	where the participants in this story share their impressions of this 
	progress. As a result, the Standard Model of unified
	interactions
	got its main technical tool.
	From the middle of the 70s until our time it remains 
	the fundamental base of high energy physics. For our discourse
	it is important to stress once again that the paper 
\cite{Faddeev}
	based on mathematical intuition preceded the works made in the 
	traditions of theoretical physics.

	The Standard Model did not complete the development of fundamental 
	physics in spite of its unexpected and astonishing experimental
	success. The gravitational interactions, whose geometrical 
	interpretation is slightly different from that of the Yang--Mills
	theory, is not included in the Standard Model. The unification of 
	quantum principles, Lorentz--Einstein relativity and Einstein
	gravity has not yet been accomplished. We have every reason
	to conjecture that the modern MP and its mode of working will play the
	decisive role in the quest for such a unification.

	Indeed, the new generation of theoreticians in high energy 
	physics have received an incomparably higher mathematical education.
They are not 
	subject to the pressure of old authorities maintaining the purity of 
	physical thinking and/or terminology. Futhermore, many 
	professional mathematicians, tempted by the beauty of the 
	methods used by physicists, 
	moved to the position of the modern mathematical
	physics. Let use cite from the manifesto, written by P.~MacPherson
	during the organization of the Quantum Field Theory year at the
	School of Mathematics of the Institute for Advanced Study at
	Princeton:
\begin{quotation}	
	The goal is to create and convey an understanding, in terms 
	congenial to mathematicians, of some fundamental notions of 
	physics, such as quantum field theory. The emphasis will be on 
	developing the intuition stemming from functional integrals.

	One way to define the goals of the program is by negation,
	excluding certain important subjects commonly pursued by 
	mathematicians whose work is motivated by physics. In this 
	spirit, it is not planned to treat except peripherally the
	magnificient new applications of field theory, such as
	Seiberg-Witten equations to Donaldson theory.
	Nor is the plan to consider fundamental new constructions within 
	mathimatics that were inspired by physics, such as quantum groups 
	or vertex operator algebras. Nor is the aim to discuss how to 
	provide mathematical rigor for physical theories. Rather, the 
	goal is to develop the sort of intuition common among physicists 
	for those who are used to thought processes stemming from 
	geometry and algebra.
\end{quotation}	

	I propose to call the intuition to which MacPherson refers that of
	mathematical physics. I also recommend the reader to look at the
	instructive drawing by P.~Dijkgraaf on the dust cover of the volumes 
	of lectures given at the School
\cite{Dijkgraaf}.

        The union of these two groups constitutes an enormous intellectual
	force. In the next century we will learn if this force is capable
	of substituting for the traditional experimental base of the
development
	of fundamental physics and pertinent physical intuition.

\end{document}